\documentclass[showpacs,twocolumn,aps,preprintnumbers,letterpaper]{revtex4}
\usepackage{amsmath,amssymb}
\usepackage{epsfig}
\usepackage{graphicx}
\usepackage{amsmath}
\usepackage{slashed}
\usepackage{amsfonts}
\usepackage{epstopdf}

\newcommand{\be}{\begin{equation}}
\newcommand{\ee}{\end{equation}}
\newcommand{\bear}{\begin{eqnarray}}
\newcommand{\eear}{\end{eqnarray}}
\newcommand{\ba}{\begin{array}}
\newcommand{\ea}{\end{array}}

\def\be{\begin{eqnarray}}
\def\ee{\end{eqnarray}}
\def\bea{\be}
\def\eea{\ee}

\def\roughly#1{\mathrel{\raise.3ex\hbox{$#1$\kern-.75em%
\lower1ex\hbox{$\sim$}}}}

\begin{document}

\title{Heavy and Strange Holographic Baryons}

\author{Yizhuang Liu}
\email{yizhuang.liu@stonybrook.edu}
\affiliation{Department of Physics and Astronomy, Stony Brook University, Stony Brook, New York 11794-3800, USA}

\author{Ismail Zahed}
\email{ismail.zahed@stonybrook.edu}
\affiliation{Department of Physics and Astronomy, Stony Brook University, Stony Brook, New York 11794-3800, USA}


\date{\today}
\begin{abstract}
We extend the $D4$-$D8$ holographic construction to include three chiral and one heavy flavor, to describe
heavy-light baryons with strangeness and their exotics.  At strong coupling, the heavy meson always binds to the bulk
instanton  in the form of a flavor zero mode in the fundamental representation. We quantize the ensuing bound states
using the collective quantization method, to obtain the spectra of heavy and strange baryons with both explicit
and hidden  charm and bottom. Our results confirm the existence of two low-lying charmed penta-quark states with 
$\frac 12^-,\frac 32^-$ assignments,  and predict many new ones with both charm and bottom. They also 
suggest a quartet of low-lying neutral $\Omega_c^0$ with assignments
$\frac 12^\pm ,\frac 32^\pm$ that are  heavier than the quintuplet of 
neutral $\Omega_c^0$ recently reported by LHCb.
\end{abstract}



\maketitle

\setcounter{footnote}{0}


\section{Introduction}

Recently the Belle collaboration~\cite{BELLE} and the BESIII collaboration~\cite{BESIII} have reported many
multiquark exotics uncommensurate with quarkonia, e.g.  the neutral $X(3872)$ and the charged $Z_c(3900)^\pm$ 
and $Z_b(10610)^\pm$.  These exotics have been also confirmed 
 by the DO collaboration at Fermilab~\cite{DO},  and the  LHCb collaboration at CERN~\cite{LHCb}. 
Also recently, the same LHCb collaboration has reported new pentaquark states $P_c^+(4380)$ and $P_c^+(4450)$ through the
decays $\Lambda_b^0\rightarrow J\Psi pK^-, J\Psi p\pi^-$~\cite{LHCbx}, and five narrow and neutral excited $\Omega^0_c$
baryon states that decay primarily to  $\Xi_c^+K^-$~\cite{LHCbxx}.
These flurry of experimental results  support  new physics involving heavy-light multiquark states, a priori  outside the 
canonical classification of the quark model.

Some of the tetra-states exotics maybe understood as
molecular bound states mediated by one-pion exchange much
like deuterons or deusons~\cite{MOLECULES,THORSSON,KARLINER,OTHERS,OTHERSX,OTHERSZ,OTHERSXX,LIUMOLECULE}.
Non-molecular heavy exotics were also discussed using constituent quark models~\cite{MANOHAR}, 
heavy solitonic baryons~\cite{RISKA,MACIEK2}, instantons~\cite{MACIEK3} and QCD sum rules~\cite{SUHONG}. 
A flurry of  quark-based descriptions of the reported neutrals $\Omega_c^0$ states have also been proposed~\cite{OMEGAC}
following earlier descriptions~\cite{EARLIER}, including a recent lattice simulation~\cite{LATTICEC}. 

The penta-states exotics reported in~\cite{LHCbx} have been foreseen in~\cite{MAREK} and since addressed by many
using both molecular and diquark constructions~\cite{MANY},  as well as a bound anti-charm to a Skyrmion~\cite{PENTARHO}.
String based pictures using string junctions~\cite{VENEZIANO} have also been suggested for the description of exotics,
including a recent proposal  in the context of the holographic inspired string hadron model~\cite{COBI}.

In QCD the light quark sector (u, d, s) is dominated by the spontaneous breaking of chiral
symmetry, while the heavy quark sector  (c, b, t) exhibits heavy-quark symmetry~\cite{ISGUR}. 
Both symmetries are at the origin of the chiral doubling in heavy-light mesons~\cite{MACIEK,BARDEEN}, 
as measured by both the BaBar collaboration~\cite{BABAR}  and the CLEOII collaboration~\cite{CLEOII}.
As most of the heavy hadrons and their exotics exhibit radiative decays through light or heavy-light mesons
it is important to formulate a non-perturbative model of QCD that honors both chiral and heavy quark
symmetry.

The initial holographic construction offers a framework for addressing  chiral symmetry and confinement
in the  double limit of large $N_c$ and large t$^\prime$Hooft coupling $\lambda=g^2N_c$. A concrete 
model was proposed by Sakai and Sugimoto~\cite{SSX}  using a $D4$-$D8$ brane construction. The
induced gravity on the probe $N_f$ $D8$ branes due to the large stack of $N_c$ $D4$ branes, causes
the probe branes to fuse in the holographic direction, providing a geometrical mechanism for the spontaneous
breaking of chiral symmetry.  The Dirac-Born-Infeld ( DBI) action on the probe branes yields a low-energy effective
action for the light pseudoscalars with full global chiral symmetry, where the vectors and axial-vector light mesons 
are dynamical gauge particles of a hidden chiral symmetry~\cite{HIDDEN}.  This construction was recently
extended to accomodate heavy mesons with explicit heavy quark symmetry~\cite{LIUHEAVY}.
The construction makes use use of  an additional heavy probe $D8$ brane in bulk~\cite{LIUHEAVY}.

In the $D4$-$D8$ brane construction, 
baryons are identified  with small size instantons by wrapping $D4$ around $S^4$,  and are dual to Skyrmions on 
the boundary~\cite{SSXB,SSXBB}. Remarkably, this identification provides a geometrical description of the baryonic core that
is so elusive in most Skyrme models~\cite{SKYRME}. A first principle description of the baryonic core is paramount to the understanding
of heavy hadrons and their exotics since the heavy quarks bind over their small Compton wavelength. 
In a  recent analysis we have shown how  heavy baryons and their exotics can be
derived from the zero modes of bulk instantons using two light flavors~\cite{LIUBARYON}.
This paper extends this analysis to the case of three light and one heavy flavors with both chiral
and heavy quark symmetry.  A key new feature of the three flavor case is the subtle form of the 
Chern-Simons term~\cite{CSLIGHT,CSTHREE} and its importance in fixing the baryonic  hypercharge in the
presence of a heavy flavor. The effect of the light strange quark mass is introduced using a bulk instanton
holonomy and treated perturbatively~\cite{WORLD}. The model allows for the description of heavy baryons as 
a bound instanton zero mode in the double limit of strong coupling followed by a large
heavy quark mass. 
This approach
will extend the bound state approach developed in the context of the Skyrme model with heavy mesons~\cite{PENTARHO,SKYRMEHEAVY} 
to holography. We note that alternative holographic models for the description of heavy hadrons have been
developed  in~\cite{FEWX,BRODSKY} without the dual strictures of chiral and heavy quark symmetrty.

The organization of the paper is as follows: In section 2 and 3 we 
briefly recall  the geometrical set up  for the derivation of the heavy-light  effective action  
for three flavors in terms of the  bulk DBI and CS actions. 
We detail the heavy-meson interactions to the flavor instanton,
and the ensuing heavy meson bound state to the instanton in bulk in the double limit
of large coupling and heavy meson  mass.  In section 4 and 5,  we use the collective quantization 
approach to derive the pertinent spectra for holographic heavy baryons and their exotics with strangeness. 
Our conclusions are in section 6. In the Appendix we briefly review the collective quantization
of the light baryons for  $N_f=2,3$.

\section{ Holographic effective action}

\subsection{D-brane set up}

\begin{figure}[h!]
\begin{center}
 \includegraphics[width=6cm]{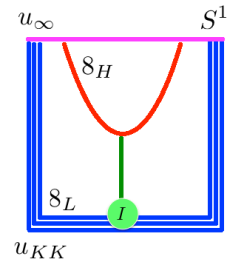}
  \caption{$N_f-1=3$ antipodal $8_L$ light branes, and one $8_H$  heavy brane shown in the $\tau U$ plane,
  with a bulk $SU(3)$ instanton embedded in $8_L$ and a massive $HL$-string connecting them.}
 \label{fig_branex}
 \end{center}
\end{figure}

The $D4$-$D8$ construction proposed by Sakai and Sugimoto~\cite{SSX}
for the description of the light hadrons  is standard and will not be repeated here. Instead, we follow~\cite{LIUHEAVY} and
consider the variant with  $N_f-1$ light $D8$-$\bar D8$ (L) and one heavy (H) probe branes in the cigar-shaped geometry that
spontaneously breaks chiral symmetry.  A schematic description  of the set up for $N_f=3$ is shown in Fig.~\ref{fig_branex}.
We assume that the L-brane world volume consists of $R^{4}\times S^1\times S^4$ with 
$[0-9]$-dimensions.  The light 8-branes are embedded in the $[0-3+5-9]$-dimensions and set
at the antipodes of $S^1$ which lies in the 4-dimension. 
The warped $[5-9]$-space is characterized by a finite size  $R$ and a horizon at $U_{KK}$.

\subsection{DBI action}


The effective action on the probe L-branes
consists of the non-Abelian DBI   and CS  action.  
After integrating over the $S^4$, the leading contribution in $1/\lambda$ to the DBI action is

\bea
\label{1}
S_{\rm DBI}\approx -\kappa\int d^4x dz\,{\rm Tr}\left(f(z){\bf F}_{\mu\nu}{\bf F}^{\mu\nu}+g(z){\bf F}_{\mu z}{\bf F}^{\nu z}\right)
\eea
Our conventions are $(-1,1,1,1)$ with $A_{M}^{\dagger}=-A_M$. The warping factors are 

\be
f(z)=\frac{R^3}{4U_z}\,,\qquad g(z)=\frac{9}{8}\frac{U_z^3}{U_{KK}}
\ee 
with $U_z^3=U_{KK}^3+U_{KK}z^2$, and $\kappa\equiv a\lambda N_c$ and
$a=1/(216\pi^3)$~\cite{SSX}.
The effective fields in the field strengths are
($M,N$ run over $(\mu,z)$)

\bea
\label{2}
&&{\bf F}_{MN}=\nonumber \\ 
&&\left(\begin{array}{cc}
F_{MN}-\Phi_{[M}\Phi_{N]}^{\dagger}&\partial_{[M}\Phi_{N]}+A_{[M}\Phi_{N]}\\
-\partial_{[M}\Phi^{\dagger}_{N]}-\Phi^{\dagger}_{[M}A_{N]}&-\Phi^{\dagger}_{[M}\Phi_{N]}
\end{array}\right)
\eea
The  matrix valued 1-form gauge field is
\be
\label{7}
{\bf A}=\left(\begin{array}{cc}
A&\Phi\\
-\Phi^{\dagger}&0
\end{array}\right)
\ee

For $N_f$ coincidental branes, the $\Phi$ multiplet is massless. However, their brane world-volume 
supports an adjoint and traceless scalar $\Psi$ in addition to the adjoint gauge field $A_M$,
which we have omitted from the DBI action for notational simplicity. The scalar admits a quartic potential 
with finite extrema and a vev $v$ for the diagonal of $\Psi$~\cite{MEYERS}, leading to a Higgs-like mass
for  the $\Phi$ multiplet 

\bea
\label{8X3}
\frac 12 m_H^2 {\rm Tr}\left(\Phi^\dagger_M \Phi_M\right)\sim \frac 12 v^2{\rm Tr}\left(\Phi^\dagger_M \Phi_M\right)
\eea
The vev is related to the separation between the light and heavy branes~\cite{MEYERS}, 
which is about the length of  the HL string. Below, $m_H$ will be taken as
the heavy meson mass for the heavy-light $(0^-,1^-)$, i.e.  $(D, D^*)$ for
charm and $(B, B^*)$ for bottom.
The introduction of a finite non-zero strange quark mass will be discussed
also below.

\subsection{Chern-Simons action}

For $N_f>2$, the naive Chern-Simons 5-form 

\be
\label{CSNAIVE}
S_{CS}=\frac{iN_c}{24\pi^2}\int_{M_5}\,{\rm Tr}\left(AF^2-\frac{1}{2}A^3F+\frac{1}{10}A^5\right)
\ee
fails to reproduce the correct transformation law under the combined gauge and chiral transformations~\cite{CSLIGHT}.
In particular, when addressing the $N_f=3$ baryon spectra, (\ref{CSNAIVE}) fails to reproduce the important hypercharge
constraint~\cite{CSLIGHT}

\be
J_8=\frac{N_c}{2\sqrt{3}}
\ee
This issue was recently revisited in~\cite{CSTHREE} where boundary contributions were added to
(\ref{CSNAIVE}) to address these shortcomings. Specifically, the new Chern-Simons (nCS) contribution is~\cite{CSTHREE}

\bea
\label{CSNEW}
&&S_{nCS}=S_{CS}\nonumber\\
&&+\int _{N_5}\frac{1}{10}\,{\rm Tr}\left(h^{-1}dh\right)^5+\int_{\partial M_5}\alpha_4\left(dh h^{-1},A\right)
\eea
Here $N_5$ is a 5-dimensional manidold whose boundaries are $\partial N_5= \partial M_5 =M_{4+\infty}-M_{4-\infty}$,
with the asymptotic flavor gauge field 

\be
A|_{z\rightarrow\pm \infty}=\hat A^{\pm h^{\pm}}=h^{\pm}(d+\hat A^{\pm})h^{\pm-1}
\ee
The gauged 4-form $\alpha_4$ is given in~\cite{CSTHREE}.
$\hat A^{\pm}$ refer to the external gauge fields, and  $h|_{\partial M_5}=(h^{+},h^{-})$. A is assumed to be well defined 
throughout  $M_5$ and produces no-boundary  contributions. In other words, in this gauge all topological information is moved to the holographic  boundaries at $z=\pm \infty$. We can actually work in the $A_z=0$ gauge, and for the instanton profile 
(as discussed below)  we have

\be
(h^{-},h^{+})\equiv \left(1,Pe^{-\int_{-\infty}^{\infty}A_zdz }\right)
\ee
Note that in our case $A\rightarrow {\bf A}$ as defined in (\ref{7}). As a result, the contributions from (\ref{CSNAIVE})
are similar to those in the  $N_f=2$ case discussed in~\cite{LIUHEAVY}. 
The contributions from the new terms in (\ref{CSNEW}) will be detailed
in the quantization approach below.

\section{Heavy-Light Baryons}

\subsection{Bulk instanton}

In the original Sakai and Sugimoto model~\cite{SSX} light baryons  are identified with small size flavor instantons in bulk~\cite{SSXB}.
This construction carries to our current set up as we have recently shown for the $N_f=2$ case in~\cite{LIUHEAVY}. 
For the present $N_f=3$ case shown in Fig.~\ref{fig_branex}, a small
size instanton translates to a flat space 4-dimensional  instanton in the $[1-4]$ directions. 
Specifically,  the SU(3) flavor instanton $A_M$ and its time components are ~\cite{CSLIGHT}

\begin{eqnarray}
\label{INST}
A_{M}=&&{\rm diag}\left(-\bar\sigma_{MN}\frac{x_N}{x^2+\rho^2},0\right)\\
A_0=&&\frac{-1}{8\pi^2a x^2}\sqrt{\frac{2}{3}}\left(1-\frac{\rho^2}{(x^2+\rho^2)^2}\right){\rm diag} (1,1,0)\nonumber \\ 
&&+\frac{1}{16\pi^2a x^2}\left(1-\frac{\rho^2}{(x^2+\rho^2)^2}\right)
{\rm diag}\left(\frac{1}{3},\frac{1}{3},-\frac{2}{3}\right)\nonumber
\end{eqnarray}
where the rescaling

\begin{eqnarray}
\label{RESCALE}
&&x_0\rightarrow x_0, x_{M}\rightarrow x_{M}/\sqrt{\lambda}, \sqrt{\lambda}\rho\rightarrow \rho\nonumber\\
&&(A_{0},\Phi_{0})\rightarrow (A_0,\Phi_0), \nonumber\\
&&(A_{M},\Phi_M)\rightarrow \sqrt{\lambda}(A_M,\Phi_M)
\end{eqnarray}
was used. From here on $M,N$ runs only over $1,2,3,z$ unless specified otherwise.
The instanton solution $A_M$ in (\ref{INST}) carries a field strength

\begin{eqnarray}
\label{INSTX}
F_{MN}={\rm diag}\left(2\frac{\bar\sigma_{MN} \rho^2}{(x^2+\rho^2)^2},0\right)
\end{eqnarray}


\subsection{Heavy-light effective action}

To order $\lambda^0$ the rescaled contributions describing the interactions between the light gauge fields $A_M$ and
the heavy fields $\Phi_M$  to quadratic order split to several contributions

\be
\label{RS1}
{\cal L}=aN_c\lambda {\cal L}_{0}+aN_c{\cal L}_{1}+{\cal L}_{CS}
\ee
with each contribution given by

\begin{eqnarray}
\label{RX0}
{\cal L}_0=&&\nonumber -(D_{M}\Phi_{N}^{\dagger}-D_{N}\Phi_{M}^{\dagger})(D_M\Phi_N-D_N\Phi_M)\nonumber \\  
&&+2\Phi_{M}^{\dagger}F_{MN}\Phi_{N}\nonumber\\
{\cal L}_1=&&+2(D_{0}\Phi_{M}^{\dagger}-D_{M}\Phi_{0}^{\dagger})(D_0\Phi_M-D_M\Phi_0)\nonumber \\ 
&&-2\Phi_0^{\dagger}F^{0M}\Phi_{M}-2\Phi_{M}^{\dagger}F^{M0}\Phi_0\nonumber\\
&&-2m_H^2\Phi_{M}^{\dagger}\Phi_M +\tilde S_1\nonumber \\
{\cal L}_{CS}=&&-\frac{iN_c}{24\pi^2}(d\Phi^{\dagger}Ad\Phi+d\Phi^{\dagger}dA\Phi+\Phi^{\dagger}dAd\Phi)\nonumber \\
&&-\frac{iN_c}{16\pi^2}(d\Phi^{\dagger} A^2\Phi+\Phi^{\dagger}A^2d\Phi+\Phi^{\dagger}(AdA+dAA)\Phi)\nonumber \\
&&-\frac{5iN_c}{48\pi^2}\Phi^{\dagger}A^3\Phi+S_C(\Phi^4,A)
\end{eqnarray}
 and 

\begin{eqnarray}
\tilde {\cal L}_1=&&+\frac{1}{3}z^2(D_i\Phi_j-D_j\Phi_i)^{\dagger}(D_i\Phi_j-D_j\Phi_i)\nonumber \\
&&-2z^2(D_i\Phi_z-D_z\Phi_i)^{\dagger}(D_i\Phi_z-D_z\Phi_i)\nonumber\\
&&-\frac{2}{3}z^2\Phi_i^{\dagger}F_{ij}\Phi_j+2z^2(\Phi_z^{\dagger}F_{zi}\Phi_i+{\rm c.c.})
\end{eqnarray}
The additional boundary contributions in (\ref{CSNEW}) do not generate any new
heavy meson contribution besides those generated by the standard Chern-Simons
contributions quoted in (\ref{RX0}).


\subsection{$\Phi$ equation of motion}

We now consider the bound state solution of the heavy meson field $\Phi_M$ in the (rescaled) instanton background
{\ref{INST}). We note that  the field equation for $\Phi_M$ is independent of $\Phi_0$ and reads

\be
\label{RX2}
D_MD_M\Phi_N+2F_{NM}\Phi_M-D_{N}D_M\Phi_M=0
\ee
while the (contraint) field equation for $\Phi_0$  depends on  $\Phi_M$ through the Chern-Simons term 

\begin{eqnarray}
\label{RX3}
&&D_M(D_0\Phi_M-D_M\Phi_0)\nonumber\\&&-F^{0M}\Phi_M-\frac{\epsilon_{MNPQ}}{64\pi^2a}K_{MNPQ}=0
\end{eqnarray}
with $K_{MNPQ}$ defined as 

 \begin{eqnarray}
 \label{RX4}
 K_{MNPQ}=&&+\partial_{M}A_N\partial_P\Phi_Q+A_MA_N\partial_P\Phi_Q\nonumber\\
 &&+\partial_MA_NA_P\Phi_Q+\frac{5}{6}A_{M}A_NA_P\Phi_Q
 \end{eqnarray}

\subsection{Heavy meson  limit}

In the heavy meson mass limit it is best to redefine 
$\Phi_{M}=\phi_{M}e^{-im_Hx_0}$ for particles. The anti-particle case follows through $m_H\rightarrow -m_H$
with pertinent sign changes. 
In the double limit of $\lambda\rightarrow \infty$ followed by $m_H\rightarrow \infty$, the leading contributions
are of order $\lambda m_H^0$ from the ligh effective action, and of order $\lambda^0m_H$ 
from the heavy-light interaction term ${\cal L}_1$ in (\ref{RX0})

\be
\label{RX5}
\frac{{\cal L}_{1,m}}{aN_c}=4im_H\phi^{\dagger}_{m}D_0\phi_{m}-2im_H(\phi_{0}^{\dagger}D_{M}\phi_{M}-{\rm c.c.})
\ee 
and the standard Chern-Simons term in (\ref{RX0})

\be
\label{RX6}
\frac{m_H N_c}{16\pi^2}\epsilon_{MNPQ}\phi^{\dagger}_{M}F_{NP}\phi_{Q}=\frac{m_HN_c}{8\pi^2}\phi^{\dagger}_{M}F_{MN}\phi_{N}
\ee
The constraint equation (\ref{RX3}) simplifies considerably to order $m_H$, that is  $D_{M}\phi_{M}=0$ and
implying that $\phi_M$ is transverse in leading order in the double limit.

\subsection{Zero-Mode}


We now observe that the heavy field equation (\ref{RX2}) in combination with the constraint
equation (\ref{RX3}) are equivalent to the vector zero-mode equation in the fundamental representation.
For  that, we recall that the field strength (\ref{INSTX}) is self-dual, and ${\cal L}_0$ in (\ref{RX0}) can be
written in the compact form

\begin{eqnarray}
\label{RX8}
{\cal L}_0=
-\frac{1}{2}\left|f_{MN}-\star f_{MN}\right|^2
\end{eqnarray}
using the Hodge $\star$ product, with $f_{MN}=\partial_{[M}\phi_{N]}+A_{[M}\phi_{N]}$. 
Therefore, the second order field equation (\ref{RX2})  can be replaced by the anti-self-dual condition (first order)
and the transversality condition   (first order),

\begin{eqnarray}
\label{RX9}
f_{MN}-\star f_{MN}=0\qquad and \qquad 
D_{M}\phi_{M}=0
\end{eqnarray}
which are equivalent to

\be
\label{RX10}
\sigma_{M}D_{M}\psi= D \psi =0 \qquad{with}\qquad
\psi=\bar \sigma_{M}\phi_{M}
\ee
The spinor zero-mode $\psi$  is unique, and its explicit matrix form reads

\be
\label{RX11}
\psi^{a}_{\alpha \beta}=\epsilon_{\alpha a}\chi_{\beta}\frac{\rho}{(x^2+\rho^2)^{\frac{3}{2}}}\qquad with \qquad a=1,2
\ee
with explicitly $\phi_M=(\bar \sigma_{M}f(x)\chi,0)$.
Here $\chi_{\alpha}$ is a constant two-component spinor. We have to understand that only the first two component of the spin-zero modes are non-zero. It can be checked explicitly that $\phi_M$
is a solution to the first order equations (\ref{RX9}).  In the presence of the instanton, the spin-1 vector field 
binds and transmutes to a spin $\frac 12$ spinor.

\section{Quantization}


The classical bound instanton-zero-mode breaks iso-rotational, rotational and translational symmetries. To quantize it, we promote
the solution to a slowly moving and rotating solution. The leading  contribution for large  $\lambda $ is purely instantonic and
its quantization is standard and can be found in~\cite{SSXBB}, so we will assume it here. The quantization of the
subleading $\lambda^0 m_H$ contribution involves the zero-mode and  for $N_f=2$ was recently addressed in~\cite{LIUBARYON}. 
Here, we will address the new elements of the quantization for $N_f=3$.

The collective quantization method proceeds by first slowly rotating and translating the instanton configuration in bulk using

\begin{eqnarray}
\label{RX14}
\Phi\rightarrow V(a_I(t)) \Phi(X_0(t),Z(t),\rho(t),\chi(t))
\end{eqnarray}
with $\Phi_0=0$. 
Here $X_0$ is the center in the 123 directions and Z is the center in the  $z$ directon. $a_I$ is the SU(3) gauge rotation moduli. 
The moduli is composed of the collective coordinates  $X_{\alpha}\equiv(X,Z,\rho)$ and by the collective SU(3)  rotation $a_I$.
The time-dependent configuration is then introduced in the heavy-light effective action described earlier and expanded in leading
order in the time-derivatives as we now detail.



\subsection{The new Chern-Simons contributions}

The additional  Chern-Simons contributions in (\ref{CSNEW}) picks up from the collectively quantized instanton
by defining

\bea
\label{HPM}
&&h^-={\rm diag}\left({a_I(t)}^{-1},1\right)\nonumber\\
&&h^{+}=h_{0}\,{\rm diag}\left({a_I(t)}^{-1},1\right)
\eea
We now note that  the field ${\bf A}$ composed of the instanton solution $A$
 plus the zero-mode solution ${\bf \Phi}$,  carries the same topological number as the field with the instanton 
solution $A$ but ${\bf \Phi}=0$. Therefore,  $h_0$ in (\ref{HPM}) can be represented by only the latter.
With this in mind,  we insert (\ref{HPM}) in the new contributions in (\ref{CSNEW}) to obtain

\be
\label{LPM1}
S_{nCS}=S_{CS}-\frac{iN_c}{48\pi^2}\int_{M^4} dt\,{\rm Tr} \left(({a_I}^{-1}\partial_{t}{a_I})(h_0^{-1}dh_0)^3 \right)
\ee 
The heavy-light contributions from $S_{CS}$ are those in (\ref{RX0}), while the new second contribution is identical
to the one obtained in the light sector~\cite{CSTHREE}

\be
\frac{N_c}{2\sqrt{3}}a^{8}
\ee
When combined to terms emerging from the heavy sector   it will give rise to the
correct hypercharge constraint as we will show next.


\subsection{Heavy contributions in leading order}

There are four contributions to order $\lambda^0m_H$ from the heavy meson sector,  namely 

\bea
\label{FOUR}
\frac{{\cal L}}{aN_c}=&&+16im_H \chi^{\dagger}\partial_t \chi f^2-16m_H\chi^{\dagger}\chi f^2\frac{2\sqrt{6}+1}{6}A_0\nonumber \\ 
&&-m_Hf^2\chi^{\dagger}\sigma_{\mu}\Phi \bar \sigma_{\mu}\chi+m_H\chi^{\dagger}\chi f^2 \frac{3}{a\pi ^2}\frac{\rho^2}{(x^2+\rho^2)^2}
\nonumber\\
\eea
The second contribution is from the $A_0$ coupling, and the third contribution simplifies for the zero-mode

\be
\chi^{\dagger}\sigma_{\mu}\Phi \bar \sigma_{\mu}\chi=a^8\frac{8\chi^{\dagger}\chi}{\sqrt{3}}
\ee
The last contribution originates from the heavy terms in naive CS term, and also simplifies using the instanton field strength and the zero-mode

\be
\label{RX18}
\frac{im_HN_c}{8\pi^2}\phi^{\dagger}_{M}F_{MN}\phi_{N}=\frac{i3m_HN_c}{\pi^2}\frac{f^2\rho^2}{(x^2+1)^2}\chi^{\dagger}\chi
\ee
In addition to the terms retained in (\ref{FOUR}) the $\chi^\dagger \chi$ coupling to the U(1) flavor  gauge field $A_0$ induces a 
Coulomb-like correction of the form $(\chi^\dagger \chi)^2$ as we have shown in~\cite{LIUBARYON}. With this in mind and after
using the rescaling $\chi \rightarrow \chi/\sqrt{4aN_cm_H}$ in (\ref{FOUR}) we obtain

\bea
\label{FOUR1}
{\cal L}= &&+{\cal L}_0[a_I, X_\alpha]\nonumber\\
&&+i\chi^{\dagger}\partial_t\chi+\frac{\eta\,\chi^{\dagger}\chi}{32\pi^2a\rho^2} -\frac{\mu\,(\chi^{\dagger}\chi)^2}{24\pi^2a N_c\rho^2 }\nonumber \\&&+a^8\frac{N_c}{2\sqrt{3}}\left(1-\frac{\chi^{\dagger}\chi}{N_c}\right)
\eea
 where the parameters $\eta, \mu$ are given by

\bea
\label{FOUR0}
\eta\equiv 2x+1\equiv \frac{2\sqrt{6}+1}{3}+1\approx 2.966\,\,\,and\,\,\,
\mu=\frac{13}{12}
\eea
Here ${\cal L}_0[a_I, X_\alpha]$ refers to the effective action density on the moduli stemming from the contribution
of the light degrees of freedom in the instanton background without the $a^8$ term~\cite{SSXB} . 

The term linear in $a^8$ in (\ref{FOUR1}) couples to
 the hypercharge $J_8=\frac{N_c}{2\sqrt{3}}(1-\frac{\chi^{\dagger}\chi}{N_c})$.
So (\ref{FOUR1}) can be seen as an action density of light and heavy degrees of freedom 
supplemented by a hypercharge constraint, namely

\bea
\label{RX20}
&&{\cal L}\rightarrow {\cal L}_0[a_I, X_\alpha]+\chi^{\dagger}i\partial_t\chi +\frac{\eta\,\chi^{\dagger}\chi}{32\pi^2a\rho^2}-\frac{\mu\,(\chi^{\dagger}\chi)^2}{24\pi^2a N_c\rho^2 }\nonumber\\
&&J^{8}=\frac{N_c}{2\sqrt{3}}\left(1-\frac{\chi^{\dagger}\chi}{N_c}\right)
\eea
From  (\ref{FOUR0}) we note that $\eta\approx 3$ and $\mu\approx 1$ which are remarkably close to the 
same parameters derived in~\cite{LIUBARYON} for the $N_f=2$ case. These terms are inertial and 
 not sensitive to the value of $N_f$.

\subsection{Heavy-light spectra}

The quantization of (\ref{RX20}) follows the same arguments as those presented in~\cite{SSXB,CSLIGHT}
for ${\cal L}_0[a_I, X_\alpha]$ as we briefly recall in the Appendix. Let $H_0$ be the Hamiltonian 
associated to ${\cal L}_0[a_I, X_\alpha]$, then the full heavy-light  Hamiltonian for (\ref{RX20})  is

\be
\label{RX21}
H=H_0[\pi_I,\pi_X, a_I, X_\alpha]-\frac{\eta\,\chi^{\dagger}\chi}{32\pi^2a\rho^2}+\frac{\mu\,(\chi^{\dagger}\chi)^2}{24\pi^2a N_c\rho^2 }
\nonumber\\
\ee
with the new quantization rule for the spinor and the hypercharge constraint

\be
\label{RX22}
&&\chi_i\chi_j^{\dagger}+\chi_j^{\dagger}\chi_i=\delta_{ij}\nonumber\\
&&J^{8}=\frac{N_c}{2\sqrt{3}}\left(1-\frac{\chi^{\dagger}\chi}{N_c}\right)
\ee
We recall that the statistics and parity of $\chi$ were fixed in~\cite{LIUBARYON}. Specifically, 
 we note the symmetry transformation

\be
\label{RX23}
\chi\rightarrow  U\chi\qquad and\qquad 
\phi_{M}\rightarrow U \Lambda_{MN}\phi_{N}
\ee
since $U^{-1}\bar \sigma_{M}U=\Lambda_{MN}\bar\sigma_{N}$. So a rotation of the spinor $\chi$ 
is equivalent to a spatial rotation of the heavy vector meson field $\phi_M$. Since $\chi$
is in the spin $\frac 12$ representation it should be quantized as a fermion. Its parity is opposite to 
that of $\phi_M$, hence positive.
With this in mind, the total spin ${\bf J}$ is given by

\be
\label{RX24}
\vec{\bf J}=-\vec{\bf I}_{SU(2)}+\vec{\bf S}_\chi\equiv -\vec{\bf I}_{SU(2)}+ \chi^\dagger\frac{\vec\tau}2\chi
\ee
Here for a general SU(3) representation, $\vec{\bf I}_{SU(2)}$ means the induced representation for 
the first three generators, $J_{1,2,3}$ as noted in the Appendix.

The spectrum of (\ref{RX21})  follows from the one discussed in~\cite{SSXB,CSLIGHT}
and recalled in the Appendix,  with  two key modifications

\be
\label{RX25}
Q\equiv \frac{N_c}{40a\pi^2}\rightarrow \frac{N_c}{40a\pi^2} \left(1-\frac {5\eta}{4N_c} \chi^\dagger \chi+\frac{5\mu(\chi^{\dagger}\chi)^2}{3N_c^2}\right)
\ee
and the change of the hypercharge as obtained in (\ref{RX22}).
The quantum states with a single bound state $N_Q=\chi^\dagger \chi=1$ and the general 
$(p,q)$ representation for SU(3) and spin $j$ are labeled by

\be
\label{RX26}
\left| N_Q,p,q,j,n_z,n_\rho\right>\,\,\,with \,\,\, IJ^\pi=\frac l2\left(\frac l2 \pm \frac 12\right)^\pi
\ee
with $n_z=0,1,2, ..$ counting the number of quanta associated to the collective motion in the
holographic direction, and $n_\rho=0,1,2,..$ counting the number of quanta associated to the
radial breathing of the instanton core, a sort of Roper-like excitations. Following~\cite{SSXB}, 
we identify the parity of the heavy  baryon bound state  as $(-1)^{n_z}$.
Using (\ref{RX25}), the mass  spectrum for the bound heavy-light states is

\begin{eqnarray}
\label{RX27}
&&M_{NQ}=M_0+N_Q\, m_H+\nonumber\\
&&\sqrt{\frac{49}{24}+\frac{{\bf K}}{3}}+\sqrt{\frac{2}{3}}(n_z+n_\rho+1) M_{KK}
\end{eqnarray}
with 

\begin{eqnarray}
{\bf  K}=&&+\frac{2N_c^2}{5}\left(1-\frac{5\eta N_Q}{4N_c}+\frac{5\mu N_Q^2}{3N_c^2}\right)
-\frac{N_c^2}{3}\left(1-\frac{N_Q}{N_c}\right)^2\nonumber \\ 
&&+\frac{4}{3}(p^2+q^2+pq+3(p+q))-2j(j+1)
\end{eqnarray}
with $M_{KK}$ the Kaluza-Klein mass and $M_0/M_{KK}=8\pi^2\kappa$ the bulk instanton mass. 
The Kaluza-Klein scale is usually set by the light meson spectrum and is fit to reproduce the rho 
mass with $M_{KK}\sim m_\rho/\sqrt{0.61}\sim 1$ GeV~\cite{SSX}. 

(\ref{RX27}) is to be contrasted with the mass spectrum for baryons with no heavy quarks or $N_Q=0$, where 
the nucleon state is idendified as $N_Q=0,l=1,n_z=n_\rho=0$ and the Delta state as $N_Q=0,l=3,n_z=n_\rho=0$~\cite{SSXB}. The radial
excitation with $n_\rho=1$ can be identified with the radial Roper excitation of the nucleon and Delta, while the holographic excitation
with $n_z=1$ can be interpreted as the odd parity excitation of the nucleon and Delta.


\subsection{Single-heavy baryons}

Since the bound zero-mode transmutes to  a spin $\frac 12$, 
the lowest heavy baryons with one heavy quark are characterized 
by $n_z,n_\rho=0,1,N_Q=1$, and $(p,q,j)=(0,1,0)$ for $\bar {\bf 3}$ and $(p,q,j)=(2,0,1)$ for $\bf 6$ . The 
$\bar {\bf 3}$-plet states  have spin and parity $\frac{1}{2}^{+}$.
We identify them with $\Lambda_{Q},\Xi_Q(\bar {\bf 3})$. The $ {\bf  6}$-plet states 
 have $J=\frac{1}{2},\frac{3}{2}$. We identify them with 
$\Sigma_Q,\Xi_Q({\bf 6}),\Omega_{Q}$ and $\Sigma_Q^{\star},\Xi_Q({\bf 6})^{\star},\Omega_{Q}^{\star}$, respectively. 
In the absence of symmetry breaking, the mass spectra are degenerate

\begin{eqnarray}
\label{RX28}
M_{\bar {\bf 3}}=&&+M_0+m_{H}+1.75M_{KK}\nonumber\\
&&+\frac {2(n_\rho+n_z)+2}{\sqrt{6}} M_{KK}\\
M_{ {\bf 6}}=&&+M_0+m_{H}+2.103M_{KK}\nonumber\\
&&+\frac {2(n_\rho+n_z)+2}{\sqrt{6}} M_{KK}
\end{eqnarray}
or equivalently

\begin{eqnarray}
\label{RX29}
&&M_{\bar {\bf 3}}-M_{p=q=1,N_{Q}=0,j=1/2}-m_{H}=-0.570\,M_{KK}\nonumber\\
&&M_{ {\bf 6}}-M_{p=q=1,N_{Q}=0,j=1/2}-m_{H}=-0.236\,M_{KK}\nonumber\\
\end{eqnarray}
with the mass splitting $M_{ {\bf 6}}-M_{\bar {\bf 3}}=0.334\,M_{KK}$.

\subsection{Double-heavy baryons: $QQ$}

While the binding of  a pair of heavy mesons with $QQ$ or  $Q\bar Q$ content  is always BPS-like to leading
order in $1/\lambda$, the Chern-Simons contribution is twice more attractive with the $QQ$ content than with
the $Q\bar Q$ content (see below), although the Coulomb induced contribution penalizes the former and not the latter. With
this in mind, heavy baryons with two heavy quarks follow the same construct with $N_Q=2$ or $\chi^\dagger\chi\rightarrow 2$
in (\ref{RX21}-\ref{RX22}) and $J^8=1/2{\sqrt{3}}$. 
As a result, the lowest heavy baryons with two bound heavy mesons
are now characterized by $n_z, n_\rho=0,1$ and $(p,q,j)=(1,0,0)$ for the flavor ${\bf 3}$-plet
 with assignment $\frac 12^+$,  which we identify as $\Xi_{QQ}$ with $u,d$ light content,
 and $\Omega_{QQ}$ with $s$ content. To this order, their degenerate masses are given by

\be
M_{\bf 3}- M_{p=q=1,N_{Q}=0,j=1/2}-2m_H =-0.844\, M_{KK}
\ee

\subsection{Double-heavy baryons: $Q\bar Q$}

For heavy baryons containing also anti-heavy quarks we note that a rerun of the preceding arguments 
using instead the reduction $\Phi_M=\phi_Me^{+im_Hx_0}$, amounts to binding an anti-heavy-light meson to 
the bulk instanton also  in the form of a zero-mode in the fundamental representation of spin, much like the 
heavy-light meson binding.  Most
of the results are unchanged except for pertinent minus signs. For instance, when binding one heavy-light
and one anti-heavy-light (\ref{RX20}) now reads

\begin{eqnarray}
\label{RX30}
{\cal L}=&&{\cal L}_0[a_I, X_\alpha]\nonumber\\
&&+\chi_Q^{\dagger}i\partial_t\chi_Q +\frac{\eta}{32\pi^2a\rho^2}\chi_Q^{\dagger}\chi_Q \nonumber\\
&&-\chi_{\bar Q}^{\dagger}i\partial_t\chi_{\bar Q} -\frac{\eta}{32\pi^2a\rho^2}\chi_{\bar Q}^{\dagger}\chi_{\bar Q}\nonumber \\ 
&&-\frac{\mu(\chi^{\dagger}_Q\chi_Q-\chi^{\dagger}_{\bar Q}\chi_{\bar Q})^2}{24\pi^2a N_c\rho^2 }
\end{eqnarray}
with the hypercharge constraint

\be
J_{8}=\frac{N_c}{2\sqrt{3}}\left(1-\frac{\chi_Q^{\dagger}\chi_Q}{N_c}+\frac{\chi_{\bar Q}^{\dagger}\chi_{\bar Q}}{N_c}\right)
\ee

The mass spectrum for baryons with $N_Q$ heavy-quarks and $N_{\bar Q}$ anti-heavy quarks is the same as in (\ref{RX27}) with the substitution $N_Q\rightarrow N_Q-N_{\bar Q}$ to the present order of the analysis or $\lambda^0m_H$.  
For $N_Q=N_{\bar Q}=1$  the hypercharge constraint is simply $J_8=\sqrt{3}/2$.
Therefore the lowest states carry $(p,q,j)=(1,1,1/2)$ and are identified  with the baryonic states in the ${\bf 8}$-plet 
representation with the $J^\pi$ assignments 
${\frac 12}^-$ and ${\frac 32}^-$, and $(p,q,j)=(3,0,3/2)$ in the ${\bf 10}$-plet representation with $J^\pi$ assignments 
(one) ${\frac 52}^-$,  (two) ${\frac 32}^-$ and (one) ${\frac 12}^-$. Their masses are given by

\begin{eqnarray}
\label{RX31}
&&M_{\bar QQ}^{{\bf 8}}=M_N+2m_{H}+\frac{2(n_z+n_{\rho})}{\sqrt{6}}M_{KK}\nonumber\\
&&M_{\bar QQ}^{{\bf 10}}=M_N+2m_{H}+0.386M_{KK}+\frac{2(n_z+n_{\rho})}{\sqrt{6}}M_{KK}\nonumber\\
\end{eqnarray}
with the mass splitting $M_{\bar QQ}^{{\bf 10}}-M_{\bar QQ}^{{\bf 8}}=0.386\,M_{KK}$.

\section{Strange quark mass correction}

To compare the previous results for single-heavy and double-heavy baryons to some of the reported 
physical spectra, we need to address the role of a finite strange quark mass. In so far, the light flavor branes
$D\bar8$-$D8$ only connect at $U_{KK}$ because of the bulk gravity induced by $D4$, thereby 
spontaneously breaking chiral symmetry.  To break explicitly chiral symmetry, say by introducing
a finite strange quark mass, an additional bulk $D6$ brane can be introduced to connect
 $D\bar 8$ to $D8$~\cite{WORLD,KOJI}.   For the $N_f=3$ case with $m_u=m_d=0$
and finite $m_s$, the worldsheet instanton in $D6$ interpolating $D\bar 8$ to $D8$,
induces an explicit light mass breaking term for the light baryons, which takes the following form
on the moduli~\cite{KOJI} 

\be
\label{MASS1}
H_{SB}= \tau \rho^3(1-D_{88}(a_I))
\ee
with $\tau\approx |m_s\left<\bar s s\right>|$. Aside from the dependence on the moduli parameter
through $\rho^3$, the explicit symmetry breaking term (\ref{MASS1}) is standard.  An estimate of $\tau$ follows from holography,
but here we will use $\tau$ as a free parameter to be adjusted below through the baryonic spectrum. (\ref{MASS1})
will be treated in perturbation theory by averaging $\rho^3$ using the radial baryonic wavefunctions
$\phi_{{n_\rho}, {\bf K}}$ discussed in the Appendix. For $n_\rho=n_z=0$, the averaged result is

\be
\left<\rho^3\right>_{n_{\rho}=0,{\bf K}}=\frac{1}{f_{\pi}^3}\left(\frac{\sqrt{6}}{4\pi^3}\right)^{\frac{3}{2}}
\frac{\Gamma\left(1+\sqrt{\frac{49}{4}+2{\bf K}}+\frac{3}{2}\right)}{\Gamma\left(1+\sqrt{\frac{49}{4}+2{\bf K}}\right)}
\ee
The emergence of the pion decay constant $f_\pi=93$ MeV follows from the holographic $\rho$-wavefunction 
as discussed in the Appendix. 
For the $\bar{\bf 3}$-plet  and ${\bf 6}$-plet  representations, we have specifically

\bea
\left<\rho^3\right>_{\bar 3}=\frac{1}{f_{\pi}^3}\left(\frac{\sqrt{6}}{4\pi^3}\right)^{\frac{3}{2}}\times 13.65\nonumber\\
\left<\rho^3\right>_{6}=\frac{1}{f_{\pi}^3}\left(\frac{\sqrt{6}}{4\pi^3}\right)^{\frac{3}{2}}\times 16.70
\eea

The corresponding mass shifts induced by the explicit symmetry breaking term (\ref{MASS1}) on the heavy-light baryonic spectra
is then

\be
\Delta M_{i}=b_i(1-a_i)\frac{\tau}{f_{\pi}^3}\left(\frac{\sqrt{6}}{4\pi^3}\right)^{\frac{3}{2}}\equiv b_i(1-a_i)m_0 
\ee
with the representation dependent parameters 

\bea
&&b_i=\frac{\Gamma(1+\sqrt{\frac{49}{4}+2{\bf K}_i}+\frac{3}{2})}{\Gamma(1+\sqrt{\frac{49}{4}+2{\bf K}_i})}\nonumber\\
&&a_i=\left<pq,j|D_{88}|pq,j\right>
\eea
For the specific representations of relevance to our analysis we have

\bea
&&a_N=\frac{3}{10},\qquad  b_{N}=18.97\nonumber\\
&&a_{\Lambda}=\frac{1}{4}, \qquad \,\,\,a_{\Xi^3}=-\frac 18\nonumber\\
&&a_{\Sigma}=\frac 1{10}, \qquad a_{\Xi^6}=-\frac 1{20},\qquad a_{\Omega}=-\frac 15
\eea

\subsection{Single-heavy baryon spectrum}

Combining all the previous results for the heavy-light masses, including the 
correction induced by the strange quark mass symmetry breaking term
(\ref{MASS1}) yield the following mass spectrum for the single-heavy
baryons

\bea
&&m_{\Lambda_{Q}}=m_{N}+m_{H}-0.57M_{KK}-3.04m_0\nonumber\\
&&m_{\Xi(\bar 3)_Q}=m_{N}+m_{H}-0.57M_{KK}+2.08m_0\nonumber\\
&&m_{\Sigma_Q}=m_{N}+m_H-0.236M_{KK}+1.75m_0\nonumber\\
&&m_{\Xi(6)_Q}=m_{N}+m_H-0.236M_{KK}+4.25m_0\nonumber\\
&&m_{\Omega_Q}=m_{N}+m_H-0.236M_{KK}+6.76m_0
\eea
In the original Sakai and Sugimoto analysis, the Kaluza-Klein parameter is fixed by the light rho mass
as indicated earlier with $M_{KK}\approx 1$ GeV. Although we will use this value for all the heavy-light
baryon masses to follow, we note that this value of $M_{KK}$ was noted to be large in~\cite{SSXB,CSLIGHT}. 
The nucleon mass $m_N=938$ MeV is set to its empirical
value. The symmetry breaking parameter  $m_0$ will be fitted to reproduce the mass splitting between the nucleon
in the octet and the $\Omega^-=sss$ in the decuplet as it is the baryon with the largest  strangeness. Specifically, we set

\be
m_{\Omega^-}-m_{N}=0.386\, M_{KK}+15.32\,m_0=732\,{\rm MeV}
\ee
which fixes $m_0=22.6$ MeV. 

So for $n_z=n_\rho=0$, the lowest heavy-light mass spectra corrected in first order 
by the light strange quark symmetry breaking, with their $J^\pi$ assignments are


\bea
&&\Lambda_{Q}(\frac{1}{2})^{+},M=m_{N}+m_{H}-0.57M_{KK}-3.04m_0\nonumber\\
&&\Xi_Q^{\bar 3}(\frac{1}{2})^{+},M=m_{N}+m_{H}-0.57M_{KK}+2.08m_0\nonumber\\
&&\Sigma_Q(\frac{1}{2})^{+},M=m_{N}+m_H-0.236M_{KK}+1.75m_0\nonumber\\
&&\Xi_Q^{6}(\frac{1}{2})^{+},M=m_{N}+m_H-0.236M_{KK}+4.25m_0\nonumber\\
&&\Omega_Q(\frac{1}{2})^{+},M=m_{N}+m_H-0.236M_{KK}+6.76m_0\nonumber\\
&&\Sigma_Q^{\star}(\frac{3}{2})^{+},M=m_{N}+m_H-0.236M_{KK}+1.75m_0\nonumber\\
&&\Xi_Q^{6\star}(\frac{3}{2})^{+},M=m_{N}+m_H-0.236M_{KK}+4.25m_0\nonumber\\
&&\Omega_Q^{\star}(\frac{3}{2})^{+},M=m_{N}+m_H-0.236M_{KK}+6.76m_0\nonumber\\
\ee
The lowest excited states of these heavy-light baryons  carry finite $n_\rho, n_z$.
For instance, for  $n_{\rho}=1$, $n_z=0$ we have the even-parity or Roper-like excitation corresponding to
$\Omega_{EQ}(\frac{1}{2})^{+}$, and for $n_\rho=0$ and $n_z=1$  we have the odd-parity
excitation corresponding to $\Omega_{Q}(\frac{1}{2})^{-}$.  Their masses are

\bea
\Omega_{Q}(\frac{1}{2})^{-},M&=&m_{N}+m_H+0.580M_{KK}+6.76m_0\nonumber\\
\Omega_{EQ}(\frac{1}{2})^{+},M&=&m_{N}+m_H+0.580M_{KK}+10.74m_0\nonumber\\
\eea
The masses of the single-heavy light baryons with charm follow by setting the charm heavy meson
mass $m_H$ to its empirical value $m_H=m_D=1870$ MeV, and similarly  for the bottom heavy meson
mass $m_H=m_B=5279$ MeV. The specifics mass values are quoted below in [MeV] with the 
measured masses from~\cite{PDG} indicated in bold numbers.

\subsubsection{Charm baryon masses {\rm [MeV]}}

\be
\Lambda_{c}(\frac{1}{2})^{+},M&=&2117\, {\bf [2286]}\nonumber\\
\Xi_c^{\bar 3}(\frac{1}{2})^{+},M&=&2320\,  {\bf [2468]}\nonumber\\
\Sigma_c(\frac{1}{2})^{+},\Sigma_c^{\star}(\frac{3}{2})^{+},M&=&2641\, {\bf [2453,2518]} \nonumber\\
\Xi_c^{6}(\frac{1}{2})^{+},\Xi_c^{6\star}(\frac{3}{2})^{+},M&=&2740\, {\bf [2576,2646]} \nonumber\\
\Omega_c(\frac{1}{2})^{+},\Omega_c^{\star}(\frac{3}{2})^{+},M&=&2840\, {\bf [2695,2766]} \nonumber\\
\Omega_{c}(\frac{1}{2})^{-},\Omega_{c}^{\star}(\frac{3}{2})^{-},M&=&3656\,{\bf [3050,3066]}\,  \nonumber\\
\Omega_{Ec}(\frac{1}{2})^{+},\Omega_{Ec}^{\star}(\frac{3}{2})^{+},M&=&3813\,{\bf [3090,3119]}\, 
\ee


\subsubsection{Bottom baryon masses {\rm [MeV]}}

\be
\Lambda_{b}(\frac{1}{2})^{+},M&=&5580\, {\bf [5619]} \nonumber\\
\Xi_b^{\bar 3}(\frac{1}{2})^{+},M&=&5696\, {\bf [5799]} \nonumber\\
\Sigma_b(\frac{1}{2})^{+},\Sigma_b^{\star}(\frac{3}{2})^{+},M&=&6022\,{\bf [5813,5834]} \nonumber\\
\Xi_b^{6}(\frac{1}{2})^{+},\Xi_b^{6\star}(\frac{3}{2})^{+},M&=&6079\, {\bf [****, 5955]} \nonumber\\
\Omega_b(\frac{1}{2})^{+},\Omega_b^{\star}(\frac{3}{2})^{+},M&=&6153\,{\bf [6048,****]} \nonumber\\
\Omega_{b}(\frac{1}{2})^{-},\Omega_{b}^{\star}(\frac{3}{2})^{-},M&=&6951\, \nonumber\\
\Omega_{Eb}(\frac{1}{2})^{+},\Omega_{Eb}^{\star}(\frac{3}{2})^{+},M&=&7041\, 
\ee

\subsection{Double-heavy baryon spectrum}

The double-heavy baryons with hidden charm or bottom are currently referred to as pentaquarks. 
Their masses in the ${\bf 8}$-plet of the flavor representation (\ref{RX31}) corrected by the strange 
quark mass are

\be
N_{\bar QQ}^{(\frac{1}{2},\frac{3}{2})^-},\,M &=& m_N+2m_H\nonumber\\
\Lambda_{\bar QQ}^{(\frac{1}{2},\frac{3}{2})^-}, \,M &=& m_N+2m_H+3.80m_0\nonumber\\
\Sigma_{\bar QQ}^{(\frac{1}{2},\frac{3}{2})^-},\,M &=& m_N+2m_H+7.59m_0\nonumber\\
\Xi_{\bar QQ}^{(\frac{1}{2},\frac{3}{2})^-},\, M &=& m_N+2m_H+9.48m_0
\ee
The penta-quark masses in the ${\bf 10}$-plet representation corrected by the strange quark mass  are
\be
\Delta_{\bar QQ}^{(\frac{1}{2},\frac{3}{2},\frac{5}{2})^-},\,M &=& m_N+2m_H+0.386M_{KK}+6.74m_0\nonumber\\
\Sigma_{\bar QQ}^{\star(\frac{1}{2},\frac{3}{2},\frac{5}{2})^-},\, M &=& m_N+2m_H+0.386M_{KK}+9.60m_0\nonumber\\
\Xi_{\bar QQ}^{\star(\frac{1}{2},\frac{3}{2},\frac{5}{2})^-},\, M &=& m_N+2m_H+0.386M_{KK}+12.46m_0\nonumber\\
\Omega_{\bar QQ}^{(\frac{1}{2},\frac{3}{2},\frac{5}{2})^-},\,M &=& m_N+2m_H+0.386M_{KK}+15.32m_0\nonumber\\
\ee

The double heavy baryons consisting of two heavy bound mesons with explicit charm or bottom 
will be referred to by $\Xi_{QQ}$ and $\Omega_{QQ}$ in the flavor ${\bf 3}$-plet representation 
as we noted earlier. Their strangeness corrected
masses are

\bea
\Xi_{QQ}^{(\frac 12)^+}, M=m_N+2m_H-0.844M_{KK}-2.67\,m_0\nonumber \\
\Omega_{QQ}^{(\frac 12)^+}, M=m_N+2m_H-0.844M_{KK}-0.54\,m_0
\eea
It is clear, that the holographic construct also describes  their excited Roper-like with
even parity as well as their odd parity partners, which can be retrieved from our formula.

\subsubsection{Charm penta-quark masses ${\rm [MeV]}$}

\be
N_{\bar cc}(\frac{1}{2},\frac{3}{2})^-,\,M &=& 4680\,{\bf [4380,4450]}\nonumber\\
\Lambda_{\bar cc}(\frac{1}{2},\frac{3}{2})^-,\,M &=& 4766\,\nonumber\\
\Sigma_{\bar cc}(\frac{1}{2},\frac{3}{2})^-,\,M&=&4852\,\nonumber\\
\Xi_{\bar cc}(\frac{1}{2},\frac{3}{2})^-,\,M&=&4894\,\nonumber\\
\Delta_{\bar cc}(\frac{1}{2},\frac{3}{2},\frac{5}{2})^-,\,M&=&5218\,\nonumber\\
\Sigma_{\bar cc}^{\star}(\frac{1}{2},\frac{3}{2},\frac{5}{2})^-,\,M&=&5283\,\nonumber\\
\Xi_{\bar cc}^{\star}(\frac{1}{2},\frac{3}{2},\frac{5}{2})^-,\,M&=&5348\,\nonumber\\
\Omega_{\bar cc}(\frac{1}{2},\frac{3}{2},\frac{5}{2})^-,\,M&=&5412\,
\ee

\subsubsection{Mixed penta-quark masses ${\rm [MeV]}$}

\be
N_{\bar bc}(\frac{1}{2},\frac{3}{2})^-,\,M&=&8089\,\nonumber\\
\Lambda_{\bar bc}(\frac{1}{2},\frac{3}{2})^-,\,M&=&8175\,\nonumber\\
\Sigma_{\bar bc}(\frac{1}{2},\frac{3}{2})^-,\,M&=&8261\,\nonumber\\
\Xi_{\bar bc}(\frac{1}{2},\frac{3}{2})^-,\,M&=&8303\,\nonumber\\
\Delta_{\bar bc}(\frac{1}{2},\frac{3}{2},\frac{5}{2})^-,\,M&=&8627\,\nonumber\\
\Sigma_{\bar bc}^{\star}(\frac{1}{2},\frac{3}{2},\frac{5}{2})^-,\,M&=&8692\,\nonumber\\
\Xi_{\bar bc}^{\star}(\frac{1}{2},\frac{3}{2},\frac{5}{2})^-,\,M&=&8757\,\nonumber\\
\Omega_{\bar bc}(\frac{1}{2},\frac{3}{2},\frac{5}{2})^-,\,M&=&8821\,
\ee

\subsubsection{Bottom penta-quark masses ${\rm [MeV]}$}

\be
N_{\bar bb}(\frac{1}{2},\frac{3}{2})^-,\,M&=&11498\,\nonumber\\
\Lambda_{\bar bb}(\frac{1}{2},\frac{3}{2})^-,\,M&=&11583\,\nonumber\\
\Sigma_{\bar bb}(\frac{1}{2},\frac{3}{2})^-,\,M&=&11670\,\nonumber\\
\Xi_{\bar bb}(\frac{1}{2},\frac{3}{2})^-,\,M&=&11712\,\nonumber\\
\Delta_{\bar bb}(\frac{1}{2},\frac{3}{2},\frac{5}{2})^-,\,M&=&12036\,\nonumber\\
\Sigma_{\bar bb}^{\star}(\frac{1}{2},\frac{3}{2},\frac{5}{2})^-,\,M&=&12101\,\nonumber\\
\Xi_{\bar bb}^{\star}(\frac{1}{2},\frac{3}{2},\frac{5}{2})^-,\,M&=&12166\,\nonumber\\
\Omega_{\bar bb}(\frac{1}{2},\frac{3}{2},\frac{5}{2})^-,\,M&=&12230\,
\ee

\subsubsection{Charm and bottom ${\bf 3}$-plet masses ${\rm [MeV]}$}

\be
\Xi_{ cc}(\frac{1}{2})^+,\,M&=&3776\,{\bf [3519]}\nonumber\\
\Omega_{ cc}(\frac{1}{2})^+,\,M&=&3848\,\nonumber\\
\Xi_{ cb}(\frac{1}{2})^+,\,M&=&7184\,\nonumber\\
\Omega_{ cb}(\frac{1}{2})^+,\,M&=&7257\,\nonumber\\
\Xi_{ bb}(\frac{1}{2})^+,\,M&=&10584\,\nonumber\\
\Omega_{ bb}(\frac{1}{2})^+,\,M&=&10657\,
\ee

\section{Conclusions}

We have presented a top-down holographic approach to the single- and double-heavy baryons 
in the variant of $D4$-$D8$ we proposed recently~\cite{LIUHEAVY} (first reference).  To order
$\lambda m_H^0$, the heavy baryons emerge from the zero mode  after binding a heavy meson
in the multiplet $(0^-,1^-)$ to the instanton. Remarkably,  in the bulk instanton field 
 the spin 1 and odd parity heavy meson transmutes equally to a spin $\frac 12$ and  even parity 
massless fermion and anti-fermion. At subleading  order, the Chern-Simons term is attractive for 
the bound meson with a heavy quark content and repulsive  for the bound meson with heavy
anti-quark content.

One of the key differences between the $N_f=2$ and $N_f=3$ case is the role played by the
amended form of the Chern-Simons term which results in a good hypercharge quantization rule
\cite{CSLIGHT,CSTHREE}. We have shown that the rule gets modified by the presence of the 
bound zero mode states, leading to a rich heavy-light spectra for single-heavy and double-heavy baryons
with hidden charm and bottom. 
In particular, the formers follow from the $\bar{\bf 3}$ and ${\bf 6}$ flavor representations, while the latters
from the ${\bf 8}$ and ${\bf 10}$ representations for the lowest states. The holographic set up allows
for a simple description of the low-lying odd-parity and Roper-like excitations of the heavy baryons. 
Our results for $N_f=3$ with massive strangeness confirm and extend our previous findings for 
massless $N_f=2$. 

To compare our results with currently known heavy-light  charm and meson spectra,  it is necessary
to account for the light strange quark mass. In holography this is induced by a worldsheet  instanton
that connects $D8$ and $D\bar 8$~\cite{WORLD}. By accounting for this correction in leading order 
perturbation theory, we have found reasonable agreement for the lowest single-heavy baryons with
a single adjustable parameter, namely the overall strength of the symmetry breaking term.  
 The holographic model describes
2  neutral $\Omega_{c}^0, \Omega_{c}^{*0}$ states with $\frac 12^+,\frac 32^+$ assignments 
as the odd parity partners of the lowest $\Omega_{c}^0, \Omega_{c}^{*0}$ states, and 2 Roper-like
neutral states with  $\frac 12^+,\frac 32^+$  assignments as the even parity partners also of the 
lowest $\Omega_{c}^0, \Omega_{c}^{*0}$ states. The $\frac 12^-\frac 32^-$ are  predicted to be
lighter than the excited  $\frac 12^+\frac 32^+$ states, however both pairs are found to be heavier than the 
5 neutral $\Omega_c^0$ states  reported recently  by the LHCb collaboration.  

The holographic set up for the heavy baryons is remarkable by the limited number of parameters it carries. 
Once the initial parameter $\kappa$ is traded for the pion decay constant $f_\pi$,  only
the symmetry breaking parameter $m_0$  was left to be fixed in either the light or heavy sector. We choose
the latter to fix it.  Clearly, the model can and should be made more realitic through the use of  improved holographic  
QCD~\cite{KIRITSIS}.

 The shortcomings of the heavy-light holographic approach stem from the triple limits of large $N_c$, strong 
 $^\prime$t Hooft coupling $\lambda=g^2N_c$, and  heavy meson mass. The corrections in $1/m_H$ are
 straighforward but laborious and  should be studied as they shed important light on the hyperfine type splittings.
 Also, it should be useful to explore the sensitivity of our results by relaxing the value of $M_{KK}$ as fixed
 in the light meson sector and addressing the strangeness mass correction beyond leading order perturbation
 theory.  The one-meson radiative decays of the heavy baryons and their exotics can be addressed
 in this model for further comparison with the experimentally reported partial widths.

\section{Acknowledgements}

This work was supported by the U.S. Department of Energy under Contract No.
DE-FG-88ER40388.

\section{Appendix}

In this Appendix we briefly recall the key steps in the collective quantization of the
holographic light baryons for both $N_f=2,3$~\cite{SSXB,CSLIGHT,KOJI}.
For $N_f=2$ and no heavy-meson (\ref{1}) describes the light meson sector. 
 In the   large $\lambda$ limit using  the same rescaling (\ref{RESCALE}) 
to re-write the contributions of the light gauge fields, we have

\be
S=aN_c \lambda S_{YM}(A_{M},\hat A_{M})+aN_cS_1(A_0,\hat A_0,A_{M},\hat A_{M})
\ee 
Here $A$ refers to  the SU(2) part of the light gauge field, and 
$\hat A$ to its U(1) part. The equation of motion for $A_M,\hat A_M$ are at leading order of $\lambda$ 

\be
D_{N}F_{NM}=0\qquad and \qquad
\partial_N \hat F_{NM}=0
\ee
They are solved using the flat instanton 
$A_M$ and 0 for $\hat A_M$. The  equation of motion for the  time components are subleading

\begin{eqnarray}
&&D_{M}F_{0M}+\frac{1}{64\pi^2a}\epsilon_{MNPQ}\hat F_{MN}F_{PQ}=0\nonumber\\
&&\partial_{M}\hat F_{0M}+\frac{1}{64\pi^2a}\epsilon_{MNPQ}tr F_{MN}F_{PQ}=0
\end{eqnarray}

To obtain the spectrum we promote the moduli of the solution to be time dependent, i.e.

\be
({ a}_I, X_{\alpha})\rightarrow ({a}_I(t), X_{\alpha}(t))
\ee
Here ${a}_I$ refers to the moduli of the global SU(2) gauge transformation. 
In order to satisfy the constraint equation (52) (Gauss's law) we need to impose a further gauge transformation on
the field configuration 

\begin{eqnarray}
A^V_{M}=V^{\dagger}(A_{M}+\partial_{M})V\qquad and \qquad
A^V_{0}=V^{\dagger}\partial_{t}	V
\end{eqnarray}
Inserting  the transformed field configuration in the constraint equation, we find that $V$ is solved by

\begin{eqnarray}
-iV^{\dagger}\partial_{t}	V=-\partial_t X_{N}A_{N}+\chi_a [a_I]\Phi_a
\end{eqnarray}
with $\chi^a[{ a}_I]=-i{\rm Tr}(\tau^a a_I^{-1}\partial_ta_I)$.
Putting the resulting slowly moving field configuration back in the action, allows for
the light collective Hamiltonian~\cite{SSXB}

\begin{eqnarray}
&&H_0=M_0+H_{Z}+H_{\rho}\nonumber\\
&&H_{Z}=-\frac{\partial_Z^2}{2m_z}+\frac{m_z\omega_z^2}{2} Z^2\nonumber\\
&&H_{\rho}=-\frac{\nabla_y^2}{2m_y}+\frac{m_y\omega_\rho^2}{2} \rho^2+\frac{Q}{\rho^2}\nonumber\\
&&y=\rho(a_1,a_2,a_3,a_4),{a}_I=a_4+i\vec{a}\cdot\vec \tau\nonumber\\
&&m_z=\frac{m_y}{2}=8\pi^2 aN_c, \omega_z^2=\frac{2}{3}, \omega_\rho^2=\frac{1}{6}
\end{eqnarray}
So for $N_f=2$, the
 eigenstates of  $H_\rho$ are given by $T^{l}(a)R_{l,n_\rho}(\rho)$, where $T^{l}$ are the spherical harmonics on $S^3$.
Under $SO(4)=SU(2)\times SU(2)/Z_2$ they  are in the $(\frac{l}{2},\frac{l}{2})$ representations, where the two SU(2) factors are defined by the isometry ${a}_I\rightarrow V_L {a}_IV_{R}$.  The left factor is the isospin rotation, and the right factor is the  space rotation. This quantization describes ${I}=J=\frac{l}{2}$ states. The nucleon is realized as the lowest state with $l=1$ and $n_{\rho}=n_z=0$. 

For the SU(3) case  most of the analysis remains the same except for two differences:
1/ the Chern-Simons term needs amendment as explained in main text; 2/ both $A_0$ and $\hat A_0$ need
to be solved to a non-zero value at the static level as also explained in the main text. With this in mind, a general
time-dependent SU(3) rotation $a_I$ generates the new collective Hamiltonian $H_\rho$ as~\cite{CSLIGHT}

\bea
H_\rho=&&-\frac{1}{2m_y}\frac{1}{\rho^{\eta}} \partial_{\rho}(\rho^{\eta}\partial_{\rho})+\frac{1}{2}m_y\omega_\rho^2\rho^2+\frac{Q}{\rho^2}\nonumber \\
&&+\frac{2\sum_{a=1}^3J_a^2}{m_y\rho^2}+\frac{4\sum_{a=4}^{7}J_a^2}{m_y\rho^2}
\eea
We note that in holography, the inertia in the $1,2,3$ directions is twice larger than the inertia
in the $4,5,6,7$ directions reflecting on the inherent SU(2) character of the flavor instanton 
in bulk. The $J_a$ are the generators of the right representation on the
group manifold associated to $a_I$.
Given a representation $(p,q)$ and right-spin $j$, we have

\bea
&&\sum_{a=1}^{8}J_a^2=\frac{1}{3}(p^2+q^2+pq+3(p+q))\nonumber\\
&&\sum_{a=1}^3J_a^2=j(j+1)
\eea


The radial wavefunctions and energies associated to the full Hamiltonian

\be
H_0=-\frac{1}{2m_y}\frac{1}{\rho^{\eta}} \partial_{\rho}(\rho^{\eta}\partial_{\rho})+\frac{1}{2}m_y\omega_\rho^2\rho^2+\frac{\bf K}{m_y\rho^2}
\ee
are found in the form

\bea
&&\phi_{n_\rho,\rho,{\bf K}}=e^{-\frac{m_y\omega_\rho\rho^2}{2}}\rho^{\beta-\frac{\eta+1}{2}}F(-n_\rho,\beta,m_y\omega_\rho\rho^2)\nonumber\\
&&\beta=1+\left({\frac{(\eta-1)^2}{2}+2{\bf K}}\right)^{\frac 12}\nonumber\\
&&E_{n_{\rho}}=\omega_{\rho}\left(2n_\rho+1+\frac{\sqrt{(\eta-1)^2+8{\bf K}}}{2}\right)
\eea
The combination $m_y\omega_\rho\equiv16\pi^2\kappa/\sqrt{6}$ if we remember to
unwind the rescaling $\sqrt{\lambda}\rho\rightarrow\rho $ from (\ref{RESCALE}). The value 
of $\kappa$ is fixed by the pion decay constant $f_\pi^2/M_{KK}^2=\kappa/(54\pi^4)$~\cite{SSX}.
The explicit wavefunctions for the SU(3) representation with assignment
$\mu=(p,q)$ are given by

\be
|\mu,YII_3,Y_RJ_sM_s>=(-1)^{J_s-M_s}D^{\mu}_{YII,Y_RJ_sM_s}(a_I)
\ee
and the total state with one spinor attached (for a single-heavy baryon) follows by re-coupling

\bea
&&\Phi_{\mu,YII_3,Y_RJJ_3}=\nonumber \\ 
&&\sum_{h=\pm,M_s+h=J_3}C^{\frac{1}{2},J_s,J}_{h,M_s,J_3}\chi_h|\mu,YII_3,Y_RJ_sM_s>
\eea
A similar re-coupling holds for the double-heavy baryons. 
When evaluating the symmetry breaking contribution through $\left<D_{88}\right>$, 
we note that the Clebsch-Gordon coefficients play no role since they depend only on
$\mu,YI,Y_RJ_s$ and not on $M_s$.





 \vfil


\begin{thebibliography}{99} \frenchspacing







\bibitem{BELLE}
  I.~Adachi [Belle Collaboration],
  arXiv:1105.4583 [hep-ex];
  A.~Bondar {\it et al.} [Belle Collaboration],
  Phys.\ Rev.\ Lett.\  {\bf 108}, 122001 (2012)
  [arXiv:1110.2251 [hep-ex]].


\bibitem{BESIII}
  M.~Ablikim {\it et al.} [BESIII Collaboration],
  Phys.\ Rev.\ Lett.\  {\bf 110}, 252001 (2013)
  [arXiv:1303.5949 [hep-ex]].



\bibitem{DO}
  V.~M.~Abazov {\it et al.} [D0 Collaboration],
  [arXiv:1602.07588 [hep-ex]].

\bibitem{LHCb}
  R.~Aaij {\it et al.} [LHCb Collaboration],
  arXiv:1606.07895 [hep-ex];
  R.~Aaij {\it et al.} [LHCb Collaboration],
  arXiv:1606.07898 [hep-ex].
  
  
  \bibitem{LHCbx}
  R.~Aaij {\it et al.} [LHCb Collaboration],
  Phys.\ Rev.\ Lett.\  {\bf 115} (2015) 072001
  [arXiv:1507.03414 [hep-ex]];
  R.~Aaij {\it et al.} [LHCb Collaboration],
  Phys.\ Rev.\ Lett.\  {\bf 117} (2016) no.8,  082002
  [arXiv:1604.05708 [hep-ex]];
  R.~Aaij {\it et al.} [LHCb Collaboration],
  Phys.\ Rev.\ Lett.\  {\bf 117} (2016) no.8,  082003
   Addendum: [Phys.\ Rev.\ Lett.\  {\bf 117} (2016) no.10,  109902]
  [arXiv:1606.06999 [hep-ex]].


\bibitem{LHCbxx}
  R.~Aaij {\it et al.} [LHCb Collaboration],
  arXiv:1703.04639 [hep-ex].




\bibitem{MOLECULES}
  M.~B.~Voloshin and L.~B.~Okun,
  JETP Lett.\  {\bf 23}, 333 (1976)
  [Pisma Zh.\ Eksp.\ Teor.\ Fiz.\  {\bf 23}, 369 (1976)];

\bibitem{THORSSON}
  N.~A.~Tornqvist,
  Phys.\ Rev.\ Lett.\  {\bf 67}, 556 (1991);
  N.~A.~Tornqvist,
  Z.\ Phys.\ C {\bf 61}, 525 (1994)
  [hep-ph/9310247];
  N.~A.~Tornqvist,
  Phys.\ Lett.\ B {\bf 590}, 209 (2004)
  [hep-ph/0402237].



  

  
  
\bibitem{KARLINER}
  M.~Karliner and H.~J.~Lipkin,
  arXiv:0802.0649 [hep-ph];
  M.~Karliner and J.~L.~Rosner,
  Phys.\ Rev.\ Lett.\  {\bf 115} (2015) no.12,  122001
  [arXiv:1506.06386 [hep-ph]];
  M.~Karliner,
  Acta Phys.\ Polon.\ B {\bf 47}, 117 (2016).



\bibitem{OTHERS}
  C.~E.~Thomas and F.~E.~Close,
  Phys.\ Rev.\ D {\bf 78}, 034007 (2008)
  [arXiv:0805.3653 [hep-ph]];
  F.~Close, C.~Downum and C.~E.~Thomas,
  Phys.\ Rev.\ D {\bf 81}, 074033 (2010)
  [arXiv:1001.2553 [hep-ph]].


  \bibitem{OTHERSX}
  S.~Ohkoda, Y.~Yamaguchi, S.~Yasui, K.~Sudoh and A.~Hosaka,
  Phys.\ Rev.\ D {\bf 86}, 034019 (2012)
  [arXiv:1202.0760 [hep-ph]];
  S.~Ohkoda, Y.~Yamaguchi, S.~Yasui, K.~Sudoh and A.~Hosaka,
  arXiv:1209.0144 [hep-ph].


\bibitem{OTHERSZ}
  M.~T.~AlFiky, F.~Gabbiani and A.~A.~Petrov,
  Phys.\ Lett.\ B {\bf 640}, 238 (2006)
  [hep-ph/0506141];
  I.~W.~Lee, A.~Faessler, T.~Gutsche and V.~E.~Lyubovitskij,
  Phys.\ Rev.\ D {\bf 80}, 094005 (2009)
  [arXiv:0910.1009 [hep-ph]];
  M.~Suzuki,
  Phys.\ Rev.\ D {\bf 72}, 114013 (2005)
  [hep-ph/0508258];
  J.~R.~Zhang, M.~Zhong and M.~Q.~Huang,
  Phys.\ Lett.\ B {\bf 704}, 312 (2011)
  [arXiv:1105.5472 [hep-ph]];
  D.~V.~Bugg,
  Europhys.\ Lett.\  {\bf 96}, 11002 (2011)
  [arXiv:1105.5492 [hep-ph]];
  J.~Nieves and M.~P.~Valderrama,
  Phys.\ Rev.\ D {\bf 84}, 056015 (2011)
  [arXiv:1106.0600 [hep-ph]];
  M.~Cleven, F.~K.~Guo, C.~Hanhart and U.~G.~Meissner,
  Eur.\ Phys.\ J.\ A {\bf 47}, 120 (2011)
  [arXiv:1107.0254 [hep-ph]];
  T.~Mehen and J.~W.~Powell,
  Phys.\ Rev.\ D {\bf 84}, 114013 (2011)
  [arXiv:1109.3479 [hep-ph]];
  F.~K.~Guo, C.~Hidalgo-Duque, J.~Nieves and M.~P.~Valderrama,
  Phys.\ Rev.\ D {\bf 88}, 054007 (2013)
  [arXiv:1303.6608 [hep-ph]];
  Q.~Wang, C.~Hanhart and Q.~Zhao,
  Phys.\ Rev.\ Lett.\  {\bf 111}, no. 13, 132003 (2013)
  [arXiv:1303.6355 [hep-ph]];
  F.~K.~Guo, C.~Hanhart, Q.~Wang and Q.~Zhao,
  Phys.\ Rev.\ D {\bf 91} (2015) no.5,  051504
  [arXiv:1411.5584 [hep-ph]];
  X.~W.~Kang, Z.~H.~Guo and J.~A.~Oller,
  Phys.\ Rev.\ D {\bf 94} (2016) no.1,  014012
  [arXiv:1603.05546 [hep-ph]];
  X.~W.~Kang and J.~A.~Oller,
  arXiv:1612.08420 [hep-ph].


 
 \bibitem{OTHERSXX}
  E.~S.~Swanson,
  Phys.\ Rept.\  {\bf 429}, 243 (2006)
  [hep-ph/0601110];
  Z.~F.~Sun, J.~He, X.~Liu, Z.~G.~Luo and S.~L.~Zhu,
  Phys.\ Rev.\ D {\bf 84}, 054002 (2011)
  [arXiv:1106.2968 [hep-ph]];
 
 
 \bibitem{LIUMOLECULE}
  Y.~Liu and I.~Zahed,
  Phys.\ Lett.\ B {\bf 762}, 362 (2016)
  [arXiv:1608.06535 [hep-ph]];
  Y.~Liu and I.~Zahed,
  Int.\ J.\ Mod.\ Phys.\ E {\bf 26},  1740017 (2017)
  [arXiv:1610.06543 [hep-ph]];
  Y.~Liu and I.~Zahed,
  arXiv:1611.04400 [hep-ph].
 
 
\bibitem{Albaladejo:2015lob} 
  M.~Albaladejo, F.~K.~Guo, C.~Hidalgo-Duque and J.~Nieves,
  Phys.\ Lett.\ B {\bf 755}, 337 (2016)
  doi:10.1016/j.physletb.2016.02.025
  [arXiv:1512.03638 [hep-ph]].
 
 
 
 
 
\bibitem{MANOHAR}
  A.~V.~Manohar and M.~B.~Wise,
  Nucl.\ Phys.\ B {\bf 399}, 17 (1993)
  [hep-ph/9212236];
  N.~Brambilla {\it et al.},
  Eur.\ Phys.\ J.\ C {\bf 71}, 1534 (2011)
  [arXiv:1010.5827 [hep-ph]];
  M.~B.~Voloshin,
  Prog.\ Part.\ Nucl.\ Phys.\  {\bf 61}, 455 (2008)
  [arXiv:0711.4556 [hep-ph]];
  J.~M.~Richard,
  arXiv:1606.08593 [hep-ph].


\bibitem{RISKA}
  D.~O.~Riska and N.~N.~Scoccola,
  Phys.\ Lett.\ B {\bf 299}, 338 (1993).

\bibitem{MACIEK2}
  M.~A.~Nowak, I.~Zahed and M.~Rho,
  Phys.\ Lett.\ B {\bf 303}, 130 (1993).

\bibitem{MACIEK3}
  S.~Chernyshev, M.~A.~Nowak and I.~Zahed,
  Phys.\ Rev.\ D {\bf 53}, 5176 (1996)
  [hep-ph/9510326].

\bibitem{SUHONG}
  M.~Nielsen, F.~S.~Navarra and S.~H.~Lee,
  Phys.\ Rept.\  {\bf 497}, 41 (2010)
  [arXiv:0911.1958 [hep-ph]].



\bibitem{OMEGAC}
  M.~Karliner and J.~L.~Rosner,
  arXiv:1703.07774 [hep-ph];
  G.~Yang and J.~Ping,
  arXiv:1703.08845 [hep-ph];
  K.~L.~Wang, L.~Y.~Xiao, X.~H.~Zhong and Q.~Zhao,
  arXiv:1703.09130 [hep-ph];
  W.~Wang and R.~L.~Zhu,
  arXiv:1704.00179 [hep-ph];
  H.~Y.~Cheng and C.~W.~Chiang,
  arXiv:1704.00396 [hep-ph];
  H.~Huang, J.~Ping and F.~Wang,
  arXiv:1704.01421 [hep-ph];
  B.~Chen and X.~Liu,
  arXiv:1704.02583 [hep-ph];
  T.~M.~Aliev, S.~Bilmis and M.~Savci,
  arXiv:1704.03439 [hep-ph];
  H.~C.~Kim, M.~V.~Polyakov and M.~Praszalowicz,
  arXiv:1704.04082 [hep-ph].
  
    
\bibitem{LATTICEC}
  M.~Padmanath and N.~Mathur,
  arXiv:1704.00259 [hep-ph].

  
  \bibitem{EARLIER}
  D.~Ebert, R.~N.~Faustov and V.~O.~Galkin,
  Phys.\ Rev.\ D {\bf 84}, 014025 (2011)
  [arXiv:1105.0583 [hep-ph]];
  D.~Ebert, R.~N.~Faustov and V.~O.~Galkin,
  Phys.\ Lett.\ B {\bf 659}, 612 (2008)
  [arXiv:0705.2957 [hep-ph]];
  W.~Roberts and M.~Pervin,
  Int.\ J.\ Mod.\ Phys.\ A {\bf 23}, 2817 (2008)
  [arXiv:0711.2492 [nucl-th]].
  

\bibitem{MAREK}
  M.~Karliner and J.~L.~Rosner,
  Phys.\ Rev.\ Lett.\  {\bf 115}, no. 12, 122001 (2015)
  [arXiv:1506.06386 [hep-ph]];
  M.~Karliner,
  EPJ Web Conf.\  {\bf 130}, 01003 (2016).


\bibitem{MANY}
  R.~Chen, X.~Liu, X.~Q.~Li and S.~L.~Zhu,
  Phys.\ Rev.\ Lett.\  {\bf 115}, no. 13, 132002 (2015)
  [arXiv:1507.03704 [hep-ph]];
  H.~X.~Chen, W.~Chen, X.~Liu, T.~G.~Steele and S.~L.~Zhu,
  Phys.\ Rev.\ Lett.\  {\bf 115}, no. 17, 172001 (2015)
  [arXiv:1507.03717 [hep-ph]];
  L.~Roca, J.~Nieves and E.~Oset,
  Phys.\ Rev.\ D {\bf 92}, no. 9, 094003 (2015)
  [arXiv:1507.04249 [hep-ph]];
  T.~J.~Burns,
  Eur.\ Phys.\ J.\ A {\bf 51}, no. 11, 152 (2015)
  [arXiv:1509.02460 [hep-ph]].
  H.~Huang, C.~Deng, J.~Ping and F.~Wang,
  Eur.\ Phys.\ J.\ C {\bf 76}, no. 11, 624 (2016)
  [arXiv:1510.04648 [hep-ph]];
  L.~Roca and E.~Oset,
  Eur.\ Phys.\ J.\ C {\bf 76}, no. 11, 591 (2016)
  [arXiv:1602.06791 [hep-ph]];
  Q.~F.~L and Y.~B.~Dong,
  Phys.\ Rev.\ D {\bf 93}, no. 7, 074020 (2016)
  [arXiv:1603.00559 [hep-ph]];
  Y.~Shimizu, D.~Suenaga and M.~Harada,
  Phys.\ Rev.\ D {\bf 93}, no. 11, 114003 (2016)
  [arXiv:1603.02376 [hep-ph]];
  C.~W.~Shen, F.~K.~Guo, J.~J.~Xie and B.~S.~Zou,
  Nucl.\ Phys.\ A {\bf 954}, 393 (2016)
  [arXiv:1603.04672 [hep-ph]];
  M.~I.~Eides, V.~Y.~Petrov and M.~V.~Polyakov,
  Phys.\ Rev.\ D {\bf 93}, no. 5, 054039 (2016)
  [arXiv:1512.00426 [hep-ph]];
  I.~A.~Perevalova, M.~V.~Polyakov and P.~Schweitzer,
  Phys.\ Rev.\ D {\bf 94}, no. 5, 054024 (2016)
  [arXiv:1607.07008 [hep-ph]];
  V.~Kopeliovich and I.~Potashnikova,
  Phys.\ Rev.\ D {\bf 93}, no. 7, 074012 (2016);
  Y.~Yamaguchi and E.~Santopinto,
  arXiv:1606.08330 [hep-ph];
  S.~Takeuchi and M.~Takizawa,
  Phys.\ Lett.\ B {\bf 764}, 254 (2017)
  [arXiv:1608.05475 [hep-ph]].
  
  
  \bibitem{PENTARHO}
  N.~N.~Scoccola, D.~O.~Riska and M.~Rho,
  Phys.\ Rev.\ D {\bf 92}, no. 5, 051501 (2015)
  [arXiv:1508.01172 [hep-ph]].
  
  

 
 \bibitem{VENEZIANO}
  G.~Rossi and G.~Veneziano,
  JHEP {\bf 1606}, 041 (2016)
  [arXiv:1603.05830 [hep-th]].


\bibitem{COBI}
  J.~Sonnenschein and D.~Weissman,
  arXiv:1606.02732 [hep-ph].



\bibitem{ISGUR}
  E.~V.~Shuryak,
  Nucl.\ Phys.\ B {\bf 198}, 83 (1982);
  N.~Isgur and M.~B.~Wise,
  Phys.\ Rev.\ Lett.\  {\bf 66} (1991) 1130;
  A.~V.~Manohar and M.~B.~Wise,
  ``Heavy quark physics,''
  Camb.\ Monogr.\ Part.\ Phys.\ Nucl.\ Phys.\ Cosmol.\  {\bf 10}, 1 (2000).


\bibitem{MACIEK}
  M.~A.~Nowak, M.~Rho and I.~Zahed,
  Phys.\ Rev.\ D {\bf 48}, 4370 (1993)
  [hep-ph/9209272];
  M.~A.~Nowak, M.~Rho and I.~Zahed,
  Acta Phys.\ Polon.\ B {\bf 35}, 2377 (2004)
  [hep-ph/0307102].


\bibitem{BARDEEN}
  W.~A.~Bardeen and C.~T.~Hill,
  Phys.\ Rev.\ D {\bf 49} (1994) 409
  [hep-ph/9304265];
  W.~A.~Bardeen, E.~J.~Eichten and C.~T.~Hill,
  Phys.\ Rev.\ D {\bf 68}, 054024 (2003)
  [hep-ph/0305049].




   
\bibitem{BABAR}
  B.~Aubert {\it et al.} [BaBar Collaboration],
  Phys.\ Rev.\ Lett.\  {\bf 90}, 242001 (2003)
  [hep-ex/0304021].

\bibitem{CLEOII}
  D.~Besson {\it et al.} [CLEO Collaboration],
  Phys.\ Rev.\ D {\bf 68}, 032002 (2003)
  Erratum: [Phys.\ Rev.\ D {\bf 75}, 119908 (2007)]
  [hep-ex/0305100].







\bibitem{SSX}
  T.~Sakai and S.~Sugimoto,
  Prog.\ Theor.\ Phys.\  {\bf 113}, 843 (2005)
  [hep-th/0412141];
  T.~Sakai and S.~Sugimoto,
  Prog.\ Theor.\ Phys.\  {\bf 114}, 1083 (2005)
  [hep-th/0507073].


\bibitem{HIDDEN}
  T.~Fujiwara, T.~Kugo, H.~Terao, S.~Uehara and K.~Yamawaki,
  Prog.\ Theor.\ Phys.\  {\bf 73}, 926 (1985).



\bibitem{FEWX}
  A.~Paredes and P.~Talavera,
  Nucl.\ Phys.\ B {\bf 713}, 438 (2005)
  [hep-th/0412260];
  J.~Erdmenger, N.~Evans and J.~Grosse,
  JHEP {\bf 0701}, 098 (2007);
  [hep-th/0605241].
  J.~Erdmenger, K.~Ghoroku and I.~Kirsch,
  JHEP {\bf 0709} (2007) 111
  [arXiv:0706.3978 [hep-th]];
  C.~P.~Herzog, S.~A.~Stricker and A.~Vuorinen,
  JHEP {\bf 0805}, 070 (2008)
  [arXiv:0802.2956 [hep-th]];
  Y.~Bai and H.~C.~Cheng,
  JHEP {\bf 1308}, 074 (2013)
  [arXiv:1306.2944 [hep-ph]];
  K.~Hashimoto, N.~Ogawa and Y.~Yamaguchi,
  JHEP {\bf 1506}, 040 (2015)
  [arXiv:1412.5590 [hep-th]].
  J.~Sonnenschein and D.~Weissman,
  arXiv:1606.02732 [hep-ph].


\bibitem{BRODSKY}
  G.~F.~de Teramond, S.~J.~Brodsky, A.~Deur, H.~G.~Dosch and R.~S.~Sufian,
  arXiv:1611.03763 [hep-ph];
  H.~G.~Dosch, G.~F.~de Teramond and S.~J.~Brodsky,
  Phys.\ Rev.\ D {\bf 92} (2015) no.7,  074010
  [arXiv:1504.05112 [hep-ph]];
H.~G.~Dosch, G.~F.~de Teramond and S. J. Brodsky,
  Phys.\ Rev. \ D {\bf 95} (2017) no. 3, 034016
  [arXiv:1612.02370 [hep-ph]].





\bibitem{MEYERS}
  R.~C.~Myers,
  JHEP {\bf 9912}, 022 (1999)
  [hep-th/9910053].


\bibitem{SSXB}
  H.~Hata, T.~Sakai, S.~Sugimoto and S.~Yamato,
  Prog.\ Theor.\ Phys.\  {\bf 117} (2007) 1157
  [hep-th/0701280 [HEP-TH]].


\bibitem{SSXBB}
  K.~Hashimoto, T.~Sakai and S.~Sugimoto,
  Prog.\ Theor.\ Phys.\  {\bf 120} (2008) 1093
  [arXiv:0806.3122 [hep-th]];
  K.~Y.~Kim and I.~Zahed,
  JHEP {\bf 0809}, 007 (2008)
  [arXiv:0807.0033 [hep-th]].

  
  
  
  \bibitem{SKYRME}
  I.~Zahed and G.~E.~Brown,
  Phys.\ Rept.\  {\bf 142}, 1 (1986);
   Multifaceted Skyrmion, Eds. M.~Rho and I.~Zahed, World Scientific, 2016.
  
  
  
  \bibitem{LIUBARYON}
  Y.~Liu and I.~Zahed,
  arXiv:1704.03412 [hep-ph].
  
  
  \bibitem{LIUHEAVY}
  Y.~Liu and I.~Zahed,
  Phys.\ Rev.\ D {\bf 95}, no. 5, 056022 (2017)
  [arXiv:1611.03757 [hep-ph]].
  Y.~Liu and I.~Zahed,
  arXiv:1611.04400 [hep-ph].

  
  \bibitem{CSLIGHT}
  H.~Hata and M.~Murata,
  Prog.\ Theor.\ Phys.\  {\bf 119}, 461 (2008)
  [arXiv:0710.2579 [hep-th]];
  
  \bibitem{CSTHREE}
  P.~H.~C.~Lau and S.~Sugimoto,
  arXiv:1612.09503 [hep-th].
  
 
\bibitem{WORLD}
  O.~Aharony and D.~Kutasov,
  Phys.\ Rev.\ D {\bf 78}, 026005 (2008)
  [arXiv:0803.3547 [hep-th]];
  K.~Hashimoto, T.~Hirayama, F.~L.~Lin and H.~U.~Yee,
  JHEP {\bf 0807}, 089 (2008)
  [arXiv:0803.4192 [hep-th]].



 \bibitem{SKYRMEHEAVY} 
  N.~N.~Scoccola,
  Nucl.\ Phys.\ A {\bf 532}, 409C (1991);
  M.~Rho, D.~O.~Riska and N.~N.~Scoccola,
  Z.\ Phys.\ A {\bf 341}, 343 (1992);
  D.~P.~Min, Y.~s.~Oh, B.~Y.~Park and M.~Rho,
  hep-ph/9209275.
  Y.~s.~Oh, B.~Y.~Park and D.~P.~Min,
  Phys.\ Rev.\ D {\bf 49}, 4649 (1994)
  [hep-ph/9402205];
  Y.~s.~Oh, B.~Y.~Park and D.~P.~Min,
  Phys.\ Rev.\ D {\bf 50}, 3350 (1994)
  [hep-ph/9407214];
  D.~P.~Min, Y.~s.~Oh, B.~Y.~Park and M.~Rho,
  Int.\ J.\ Mod.\ Phys.\ E {\bf 4}, 47 (1995)
  [hep-ph/9412302];
  Y.~s.~Oh and B.~Y.~Park,
  Phys.\ Rev.\ D {\bf 51}, 5016 (1995)
  [hep-ph/9501356];
  J.~Schechter, A.~Subbaraman, S.~Vaidya and H.~Weigel,
  Nucl.\ Phys.\ A {\bf 590}, 655 (1995)
  Erratum: [Nucl.\ Phys.\ A {\bf 598}, 583 (1996)]
  [hep-ph/9503307];
  Y.~s.~Oh and B.~Y.~Park,
  Z.\ Phys.\ A {\bf 359}, 83 (1997)
  [hep-ph/9703219];
  C.~L.~Schat and N.~N.~Scoccola,
  Phys.\ Rev.\ D {\bf 61}, 034008 (2000)
  [hep-ph/9907271];
  N.~N.~Scoccola,
  arXiv:0905.2722 [hep-ph];
  J.~P.~Blanckenberg and H.~Weigel,
  Phys.\ Lett.\ B {\bf 750}, 230 (2015)
  [arXiv:1505.06655 [hep-ph]].


\bibitem{THETA}
  N.~Itzhaki, I.~R.~Klebanov, P.~Ouyang and L.~Rastelli,
  Nucl.\ Phys.\ B {\bf 684}, 264 (2004)
  [hep-ph/0309305].



\bibitem{HOLOXX}
  J.~M.~Maldacena,
  Int.\ J.\ Theor.\ Phys.\  {\bf 38}, 1113 (1999)
  [Adv.\ Theor.\ Math.\ Phys.\  {\bf 2}, 231 (1998)]
  [hep-th/9711200];
  S.~S.~Gubser, I.~R.~Klebanov and A.~M.~Polyakov,
  Phys.\ Lett.\ B {\bf 428}, 105 (1998)
  [hep-th/9802109];
  E.~Witten,
  Adv.\ Theor.\ Math.\ Phys.\  {\bf 2}, 505 (1998)
  [hep-th/9803131];
  I.~R.~Klebanov and E.~Witten,
  Nucl.\ Phys.\ B {\bf 556}, 89 (1999)
  [hep-th/9905104].



\bibitem{HOLOXXX}
  J.~Erlich, E.~Katz, D.~T.~Son and M.~A.~Stephanov,
  Phys.\ Rev.\ Lett.\  {\bf 95}, 261602 (2005)
  [hep-ph/0501128];
  L.~Da Rold and A.~Pomarol,
  Nucl.\ Phys.\ B {\bf 721}, 79 (2005)
  [hep-ph/0501218].





\bibitem{HOLOXXXX}
  S.~Hong, S.~Yoon and M.~J.~Strassler,
  JHEP {\bf 0604}, 003 (2006)
  [hep-th/0409118];
  J.~Erlich, G.~D.~Kribs and I.~Low,
  Phys.\ Rev.\ D {\bf 73}, 096001 (2006)
  doi:10.1103/PhysRevD.73.096001
  [hep-th/0602110];
  H.~R.~Grigoryan and A.~V.~Radyushkin,
  Phys.\ Rev.\ D {\bf 76}, 095007 (2007)
  [arXiv:0706.1543 [hep-ph]];
  H.~R.~Grigoryan and A.~V.~Radyushkin,
  Phys.\ Lett.\ B {\bf 650}, 421 (2007)
  [hep-ph/0703069];
  S.~S.~Afonin and I.~V.~Pusenkov,
  EPJ Web Conf.\  {\bf 125}, 04004 (2016)
  [arXiv:1606.06091 [hep-ph]];
  N.~R.~F.~Braga, M.~A.~Martin Contreras and S.~Diles,
  Europhys.\ Lett.\  {\bf 115}, no. 3, 31002 (2016)
  [arXiv:1511.06373 [hep-th]];
  A.~Gorsky, S.~B.~Gudnason and A.~Krikun,
  Phys.\ Rev.\ D {\bf 91}, no. 12, 126008 (2015)
  [arXiv:1503.04820 [hep-th]].














\bibitem{KOJI}
  K.~Hashimoto, N.~Iizuka, T.~Ishii and D.~Kadoh,
 Phys.\ Lett.\ B {\bf 691}, 65 (2010)
[arXiv:0910.1179 [hep-th]].

\bibitem{PDG}
C.~Patrignani et al. (Particle Data Group), Chin. Phys. {\bf C40} 100001 (2016).



\bibitem{KIRITSIS}
  U.~Gursoy and E.~Kiritsis,
  JHEP {\bf 0802}, 032 (2008)
  [arXiv:0707.1324 [hep-th]];
  U.~Gursoy, E.~Kiritsis and F.~Nitti,
  JHEP {\bf 0802}, 019 (2008)
  [arXiv:0707.1349 [hep-th]].













\end{thebibliography}
\end{document}